\documentclass[twocolumn,fp,a4paper]{jpsj3}
\usepackage{txfonts}

\usepackage{graphicx}% Include figure files
\usepackage{dcolumn}% Align table columns on decimal point
\usepackage{bm}% bold math

\usepackage{mathrsfs}
\usepackage{multirow}
\usepackage{booktabs}

\newcommand{\V}[1]{\bm{#1} } %vector command
\newcommand{\Ave}[1]{\left\langle {#1} \right\rangle} %thermal average 
\newcommand{\gsim}{\ $\raisebox{-.7ex}{$\stackrel{\textstyle >}{\sim}$}$\,\ }

\newcommand{\lb}{\left(}
\newcommand{\rb}{\right)}

\newcommand{\Req}[1]{eq.\ (\ref{eq:#1})}

\newcommand{\Rfig}[1]{Fig.\ \ref{fig:#1}}
\newcommand{\Rfigss}[2]{Figs.\ \ref{fig:#1}-\ref{fig:#2}}
\newcommand{\Lfig}[1]{\label{fig:#1}}
\newcommand{\Leq}[1]{\label{eq:#1}}
\newcommand{\Lsec}[1]{\label{sec:#1}}
\newcommand{\Rsec}[1]{\ref{sec:#1}}
\newcommand{\be}{\begin{eqnarray}}
\newcommand{\ee}{\end{eqnarray}}
\newcommand{\subbe}{\begin{subequations}}
\newcommand{\subee}{\end{subequations}}
\newcommand{\ba}{\begin{array}}
\newcommand{\ea}{\end{array}}
\newcommand{\no}{\nonumber}
\title{Spin and chiral orderings of the antiferromagnetic {\it XY\/} model on the triangular lattice and their critical properties}

\author{
Tomoyuki Obuchi\thanks{E-mail address: obuchi@spin.ess.sci.osaka-u.ac.jp} 
and 
\name{Hikaru Kawamura} 
}
\inst{\address{
Department of Earth and Space Science, Faculty of Science, Osaka University, Toyonaka, Osaka, 560-0043, Japan
}
}

\abst{We study the antiferromagnetic {\it XY\/} model on a triangular lattice by extensive Monte Carlo simulations, focusing on its ordering and critical properties. Our result clearly shows that two separate transitions occur at two distinct temperatures, the one at a higher temperature is associated with a $Z_2$-symmetry breaking driven by the chirality, and the one at a lower temperature is associated with the onset of the quasi-long-range order of the {\it XY\/} spin. We carefully examine the critical properties of each transition to find that the criticality of the chiral transition is consistent with the standard two-dimensional Ising universality class, whereas that of the spin transition might differ from the conventional Kosterlitz-Thouless (KT) one. The observed non-KT nature of the spin criticality is consistent with the most recent simulation result on the fully-frustrated {\it XY\/} model on a square lattice. 
}

\kword{frustration, triangular lattice, Monte Carlo simulation, chirality, Kosterlitz-Thouless transition}

\begin{document}

\maketitle

\section{\label{sec:intro}Introduction}

The ordering of frustrated magnets has long been studied quite extensively as one of primary subjects in statistical physics~\cite{Wannier:50,Shoji:51,Anderson:73,Villain:77,Kawamura:98}. Frustrated magnets show rich phase diagrams and new critical behaviors associated with novel physical objects which are composites of several spins~\cite{Anderson:73,Villain:77,Kawamura:98,Kawamura:10}. A typical example might be the ``chirality'', which quantifies the handedness of the noncollinear spin structure. Recent theoretical and experimental studies have revealed that the chirality often plays an essential role in the ordering of many frustrated magnets~\cite{Kawamura:98,Kawamura:10}. 

For last decades, various types of two-dimensional (2D) fully-frustrated {\it XY\/} (FFXY) models, which are believed to show a common critical behavior, have been intensively investigated~\cite{Teitel:83,Miyashita:83,D_H_Lee:84,Thijssen:90,J_Lee:91-1,J_Lee:91-2,Granato:93,S_Lee:94,Ramirez-Santiago:92,Nightingale:95,Olsson:95,Xu:96,S_Lee:98,Boubcheur:98,Ozeki:03,Hasenbusch:05,Thijssen:88,Grest:89,J_-R_Lee:94,Minnhagen:85,Minnhagen:87,Granato:05,Korshunov:02,Olsson:05,Capriotti:98,S_Lee:97,Choi:85,Granato:86,Jeon:97,Okumura:11}. They include the 2D {\it XY\/} model on a square lattice with both the ferromagnetic and the antiferromagnetic (AF) couplings \cite{Thijssen:90,J_Lee:91-2,S_Lee:94,Ramirez-Santiago:92,Olsson:95,Boubcheur:98,Ozeki:03,Hasenbusch:05,Okumura:11}, and that on a triangular lattice with the purely AF coupling \cite{Teitel:83,Miyashita:83,D_H_Lee:84,Thijssen:90,J_Lee:91-2,Xu:96,S_Lee:98,Ozeki:03}. These 2D FFXY models are known to be characterized by the $Z_2$ {\it chiral degeneracy\/}, {\it i.e.\/}, the existence of twofold degenerate ground states distinguished by the opposite signs of the chirality. The system then has a possibility to exhibit a chirality-driven phase transition associated with the spontaneous breaking of this $Z_2$ symmetry. 

 Main motivation to study the 2D FFXY models has been to reveal whether the chiral transition occurs separately from the spin transition or not. After many controversial discussions about this point in past days, recent intensive studies have established that this ``spin-chirality decoupling'' phenomenon really occurs in the 2D FFXY models. Namely, two phase transitions take place at two distinct temperatures~\cite{Korshunov:02,Ozeki:03,Hasenbusch:05,Okumura:11}. The transition at a higher temperature $T_c$ involves the $Z_2$ chiral-symmetry breaking, while the spin ordering at a lower temperature $T_s(<T_c)$ is accompanied by the onset of the quasi-long-range order of the spin similar to the Kosterlitz-Thouless (KT) transition~\cite{Kosterlitz:73}. 

 Despite the progress concerning the existence of two separate chirality and spin transitions, the critical properties of these transitions still remain somewhat controversial. For the chiral transition at $T_c$, a plausible universality class might be the 2D Ising one, since the chiral transition is accompanied by the $Z_2$-symmetry breaking. Although several studies indeed supported the Ising universality~\cite{Hasenbusch:05,Okumura:11}, some counter-evidences, either based on theoretical considerations~\cite{Jeon:97} or on numerical calculations~\cite{Grest:89,J_-R_Lee:94,S_Lee:98,Ozeki:03}, were also reported. 

 Concerning the spin transition at $T_s < T_c$, the issue of whether it belongs to the standard KT universality class or not has also been hotly debated. Indeed, possible deviations of the critical exponents from the standard KT values were reported by several researches~\cite{Teitel:83,Minnhagen:85,Choi:85,Granato:86,J_-R_Lee:94,Grest:89,S_Lee:98,Okumura:11}, while some others claimed that the spin transition belonged to the conventional KT universality class~\cite{Jeon:97,Ozeki:03,Hasenbusch:05}.

 Under these circumstances, in order to obtain further insight into the critical properties of the chiral and the spin transitions, we investigate in the present paper the ordering of the AF {\it XY\/} model on a triangular lattice by extensive Monte Carlo (MC) simulations. Among the two typical 2D FFXY models, {\it i.e.\/}, the square-lattice FFXY model and the triangular-lattice AF {\it XY\/} model, the former was recently studied quite extensively by large-scale equilibrium MC simulations up to the linear size as large as $O(10^3)$ by Hasenbusch, Pelissetto, and Vicari~\cite{Hasenbusch:05}, and by Okumura, Yoshino, and Kawamura~\cite{Okumura:11}. Concerning the latter model, {\it i.e.\/}, the triangular-lattice AF {\it XY\/} model, by contrast, a comparable numerical study has still been lacking. Equilibrium simulations were so far limited to the size of $O(10^2)$, {\it e.g.\/} up to $L=108$ in ref. 20, while nonequilibrium simulations, which dealt with an apparently large size of $L=2000$, had a limitation of the finite-time effect~\cite{Ozeki:03}. In order to fulfill this gap for the triangular-lattice AF {\it XY\/} model, and to get further information about its critical properties, we perform here a large-scale {\it equilibrium\/} MC simulation on the model up to the size $L=1152$, comparable to those recently performed for the square-lattice FFXY model.

 Our result suggests that the chiral transition belongs to the standard Ising universality, while the spin one belongs to the non-KT universality. This conclusion is fully consistent with that of the latest simulation on the FFXY model on the square lattice~\cite{Okumura:11}. 

 The remaining part of this paper is organized as follows. In \S \Rsec{model}, we introduce our model and define several physical quantities, including the specific heat, the spin and the chiral order parameters (ordering susceptibilities), the spin and the chiral Binder ratios, the spin and the chiral correlation lengths, the helicity modulus and the vorticity modulus, which are computed in our simulations and employed in our analysis of the critical properties. Details of our MC simulations are also given here. In \S \Rsec{result}, we present the result of our MC simulations. The critical properties of both the chiral and the spin transitions are analyzed in detail. Finally, \S \Rsec{summary} is devoted to summary and discussion.

\section{\label{sec:model}MODEL AND PHYSICAL QUANTITIES}
 Our model is the AF {\it XY\/} model on the triangular lattice. We assume that sites are labeled by $i,j$ ($i,j=1,2,\cdots,N$), and the corresponding coordinate is denoted as $\V{r}_{i}=(x_{i},y_{i})$. The number of spins $N$ is related to the linear system size $L$ as $N=L\times L$. The {\it XY\/} spin on a site $i$, $\V{S}_{i}$, has two components as $\V{S}_i=(S_{i}^{x},S_{i}^{y})=(\cos\theta_i,\sin\theta_i)$ where $0 \leq \theta_{i} < 2 \pi$. The Hamiltonian is given by
\be
\mathcal{H}=-\sum_{\Ave{i,j}}J_{ij}\V{S}_{i}\cdot\V{S}_j=-J\sum_{\Ave{i,j}}\cos(\theta_i-\theta_j),
\Leq{Hamiltonian}
\ee
where $J_{ij}$ is the exchange interaction and kept constant $J<0$ for the rest of the paper, and the summation $\Ave{ij}$ is taken over the all nearest neighbors. The partition function is given by
\be
e^{-\beta F}=Z=\int \prod_{i=1}^{N}\frac{d\theta_i}{2\pi} 
e^{-\beta \mathcal{H} },
\ee
where $\beta$ is the inverse temperature $1/(k_{\rm B}T)$, $k_{\rm B}$ being the Boltzmann constant. In the following, we normalize the temperature in a unit of $J/k_{\rm B}=1$. The thermal average will be denoted by the angular brackets $\Ave{\cdots}$.

\subsection{Chirality-related observables}

First, we define the local chirality on a plaquette $p$, an upward elementary triangle, as
\be
\kappa_{p}=\frac{2}{3\sqrt{3}} \sum_{\Ave{i,j}\in p}
\V{e}_z \cdot (\V{S}_{i}\times \V{S}_{j})
\no \\
=\frac{2}{3\sqrt{3}} \sum_{\Ave{i,j}\in p}\sin\lb \theta_i-\theta_j \rb,
\ee
where $\V{e}_z$ denotes a unit vector along the $z$ direction. The directed sum $\sum_{\Ave{i,j}\in p}$ is taken over three bonds surrounding a plaquette $p$ in a counter-clockwise direction. The normalization factor is chosen to satisfy the condition $|\kappa_{p}|=1$ in the ground state where the spins take the so-called $120$-degrees or the $\sqrt{3}\times \sqrt{3}$ structure. The chiralities exhibit a checkerboard pattern in the ground state. The associated chiral order parameter is defined by,
\be
m_{c}=\frac{1}{N}\sum_{j=1}^{N}\kappa_{p_{j}},
\ee
where $p_j$ denotes the index of a plaquette whose bottom-left is the site $j$. 

The chiral Binder parameter is defined by
\be
g_{c}=\frac{3}{2}\lb 1-\frac{\Ave{m_c^4}}{2\Ave{m_c^2}^2} \rb,
\ee
and the chiral correlation length is
\be
\xi_{c}=\frac{1}{2\sin(\pi/L)}\sqrt{\frac{\Ave{|m_{c}|^2}}{\Ave{|m_{c}(\V{q}_{\rm min})|^2}}-1},
\ee
where $\V{q}_{\rm min}=(2\pi/L)\V{e}_x$. We introduce the Fourier component of the chirality with wavevector $\V{q}$ by
\be
m_{c}(\V{q})=\frac{1}{N}\sum_{j=1}^{N}e^{-i\V{q} \cdot \V{r}_j}\kappa_{p_{j}}.
\ee
The chiral susceptibility is defined by
\be
\chi_{c}=N\beta \Ave{m_{c}^2}.
\ee

\subsection{Spin-related observables}

 The $\sqrt{3}\times \sqrt{3}$ spin structure can be captured via the $K$-point magnetization of the wavevector $\V{q}_{s}=(4\pi/3,0)$ 
\be
\V{m}_{s}=\frac{1}{N}\sum_{j=1}^{N}e^{-i\V{q_{s}}\cdot \V{r}_j}\V{S}_j.
\ee
The corresponding Binder parameter is defined by
\be
g_{s}=2-\frac{\Ave{|\V{m}_s|^4}}{\Ave{|\V{m}_s|^2}^2}.
\ee
The spin correlation length is given by
\be
\xi_{s}=\frac{1}{2\sin(\pi/L)}\sqrt{\frac{\Ave{|\V{m}_{s}|^2}}{\Ave{|\V{m}_{s}(\V{q}_{\rm min})|^2}}-1},
\ee
where the Fourier transform of the $K$-point magnetization is given by
\be
\V{m}_{s}(\V{q})=\frac{1}{N}\sum_{j=1}^{N}e^{-i(\V{q}_{s}+\V{q})\cdot \V{r}_{j}}
\V{S}_{j}.
\ee
 The $K$-point spin susceptibility is defined by
\be
\chi_{s}=N\beta\Ave{|\V{m}_{s}|^2}.
\ee

 In addition to the above quantities, we calculate the helicity modulus to detect the rigidity of the system against a uniaxial twist applied to the spin. This quantity is known to effectively detect the KT transition.  Consider an infinitesimal twist $\Delta \theta/L$ applied to the spin along, say, the $x$ axis.  In the Hamiltonian (\ref{eq:Hamiltonian}), we replace $(\theta_{i}-\theta_{j})$ by $(\theta_{i}-\theta_{j}+(\Delta \theta/L)\V{e}_x\cdot \V{e}_{ij})$, where $\V{e}_{ij}$ is a lattice vector connecting the site $i$ to the site $j$. Expanding the free energy with respect to $\Delta \theta$, we get
\be
\Delta F=Y'(\Delta\theta)^2+ O(\Delta\theta^4).
\ee
where  
\be
Y'=\frac{1}{N}\Biggl\{
\Ave{\sum_{\Ave{i,j}}J_{ij}(\V{e}_x\cdot \V{e}_{ij})^2 \cos(\theta_{i}-\theta_{j})}
\no \\
-\beta 
\Ave{
\lb 
\sum_{\Ave{i,j}}J_{ij}(\V{e}_x\cdot \V{e}_{ij}) \sin(\theta_{i}-\theta_{j}) 
\rb^2
}
\Biggr\}.
\ee
Since the spin density of the triangular lattice is $2/\sqrt{3}$ times larger than the square lattice~\cite{D_H_Lee:84,S_Lee:98}, we need to normalize $Y'$ to make a direct comparison with the KT theory. The normalized helicity modulus $Y$ is defined by
\be
Y=\frac{2}{\sqrt{3}}Y'.
\ee

To obtain further insight into the low-temperature phase, we also calculate the vorticity modulus~\cite{Kawamura:93}. The vorticity modulus measures the rigidity of the system against an emergence of an isolated vortex, and is originally calculated by the difference between the total free energies under the two different boundary conditions (BC): One is the usual fixed BC where the boundary spins are fixed so that they match the ground-state $\sqrt 3\times \sqrt 3$ spin configuration, and the other is the ``vortex'' BC where the boundary spins are fixed so that a single vortex is accommodated on top of the fixed-BC spin configuration~\cite{Kawamura:93}. More conveniently, the same quantity can be estimated by calculating appropriate fluctuations under the standard periodic BC~\cite{Xu:96,Okumura:11}, which we adopt in our present computation. Consider the phase shifts of spins caused by an emergence of an isolated vortex with a winding number $m$. The phase shifts are written as
\be
(\theta_{i}-\theta_{j}) \to (\theta_{i}-\theta_{j}+m \phi_{ij})
\ee
where $\phi_{ij}=-\phi_{ji}$ is an angle between the sites $i$ and $j$ with respect to the vortex core. We assume that the vortex core is located at the center of the lattice. Conventionally, the solid angle is approximated by $\phi_{ij}=\V{e}_{ij}\cdot \V{t}_{i}/r^{(c)}_{i}$ where $r^{(c)}_{i}$ is the distance between site $i$ and the vortex core, and $\V{t}_{i}$ is a unit vector  enclosing the vortex core and tangential to the circular path passing through the site $i$. Neglecting the discreteness of the vortex winding number $(m=1,2,\cdots)$ and expanding the free energy with respect to $m$, we get the vorticity modulus $V$ as
\be
\Delta F=Vm^2+O(m^4),
\ee
where
\be
V=\Ave{
\sum_{\Ave{i,j}} J_{ij}\phi_{ij}^2\cos(\theta_i-\theta_{j})
}
\no\\
-\beta \Ave{
\lb \sum_{\Ave{i,j}} J_{ij}\phi_{ij}\sin(\theta_i-\theta_j) \rb^2
}.\Leq{vorticity}
\ee
For large enough systems, the vorticity modulus is expected to behave as
\be
V(T,L)\approx c(T)+v(T)\ln L,
\ee
where we call $v(T)$ the reduced vorticity modulus. 

We expect that the present system gets the rigidity against a single vortex excitation at low temperatures, but loses it at high temperatures, {\it i.e.\/}, $v(T)$ is expected to take finite values at and below the spin transition temperature. This property might be utilized in estimating the spin transition temperature. In practice, we estimate $v(T)$ by taking the difference between the vorticity modulus of different sizes as
\be
v(T|L_1,L_2)=\frac{V(T,L_1)-V(T,L_2)}{\ln(L_1/L_2)},
\ee
for large $L_1$ and $L_2$. As an additional remark, we comment on the problem associated with the BC employed in computing the vorticity modulus. Under the periodic BC employed in the present simulation, $\phi_{ij}$ becomes ill-defined at the boundary. We avoid this problem by setting $\phi_{ij}$ to be zero in computing the vorticity modulus if the bond between $i$ and $j$ passes across the boundary.

\subsection{Details of simulations}

The standard MC method with the Metropolis update combined with the over-relaxation update is used in our simulation. Our unit MC step per spin (MCS) consists of one Metropolis sweep followed by three consecutive over-relaxation sweeps. System sizes are $L=36,72,144,288,576,1152$, and $L=48,96,192,384,768$. Typical number of total MCSs are $4\times 10^6$ MCS for $L \leq 288$, $4\times 10^7$ MCS for $384\leq L \leq 576$, and $10^8$ MCS for $768 \leq L \leq 1152$. The initial $3\times 10^6$ MCS are discarded for equilibration. The BC is taken to be periodic. We extrapolate the data at the measuring temperature to nearby temperatures by using the histogram method. This enables us to calculate physical quantities in a wide temperature range by performing a small number of simulations, which is particularly useful in precisely locating the critical temperature.

\section{\Lsec{result}MONTE CARLO RESULTS}

In this section, we present the result of our MC simulations, and examine the critical properties of the chiral and the spin transitions.

\subsection{Chiral transition}

We first present in \Rfig{C} the temperature dependence of the specific heat per spin, $C(T)=\beta^2 \lb \Ave{\mathcal{H}^2}-\Ave{\mathcal{H}}^2 \rb/N$.  The data show a clear peak around the chiral transition temperature $T=T_c$. The size dependence of the peak temperatures $T_{\rm peak}(L)$ is shown in \Rfig{peakC}. We see from the figure that there is a characteristic size around $L_{\times} \approx 300$ at which the size dependence of the peak location changes drastically. For smaller sizes $L<L_{\times}$, a near-linear dependence is observed, while, for larger sizes $L>L_{\times}$, another linear dependence with a different slope is observed. Similar crossover behavior was observed also for the square-lattice FFXY model~\cite{Okumura:11}. This crossover may be understood by considering the relation between the chiral and the spin correlation lengths around $T_c$.
\begin{figure}[tbp]
\includegraphics[width=1\columnwidth,height=0.7\columnwidth]{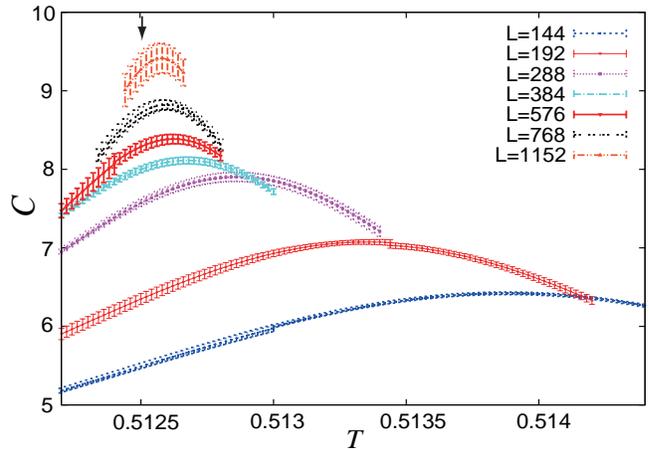}
\caption{\Lfig{C}(Color online) The temperature dependence of the specific heat per spin around the chiral transition temperature $T_{c}$. The arrow indicates our estimate of the transition temperature $T_c$. 
}
\end{figure}

\begin{figure}[tbp]
\includegraphics[width=1\columnwidth,height=0.7\columnwidth]{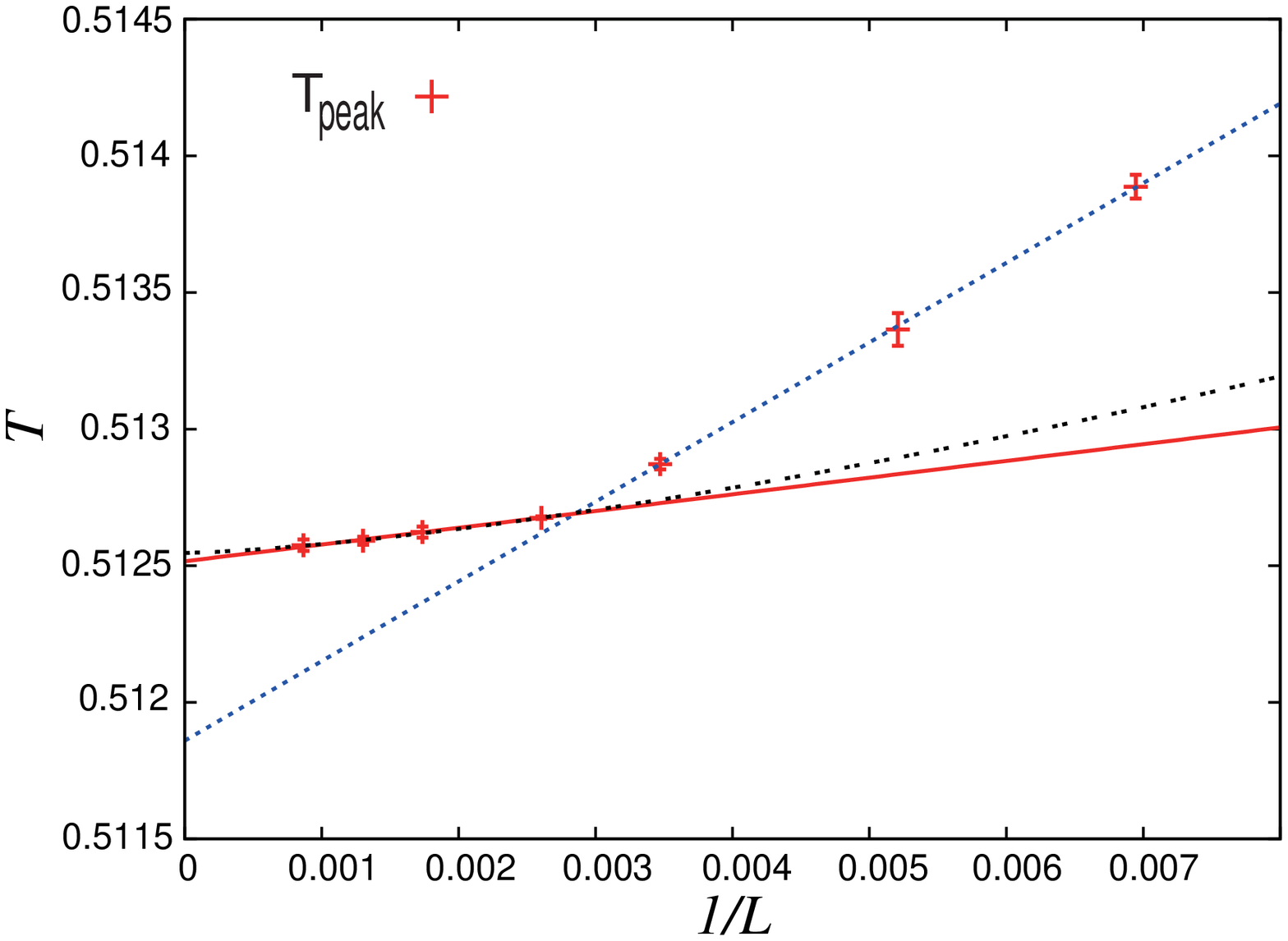}
\includegraphics[width=1\columnwidth,height=0.7\columnwidth]{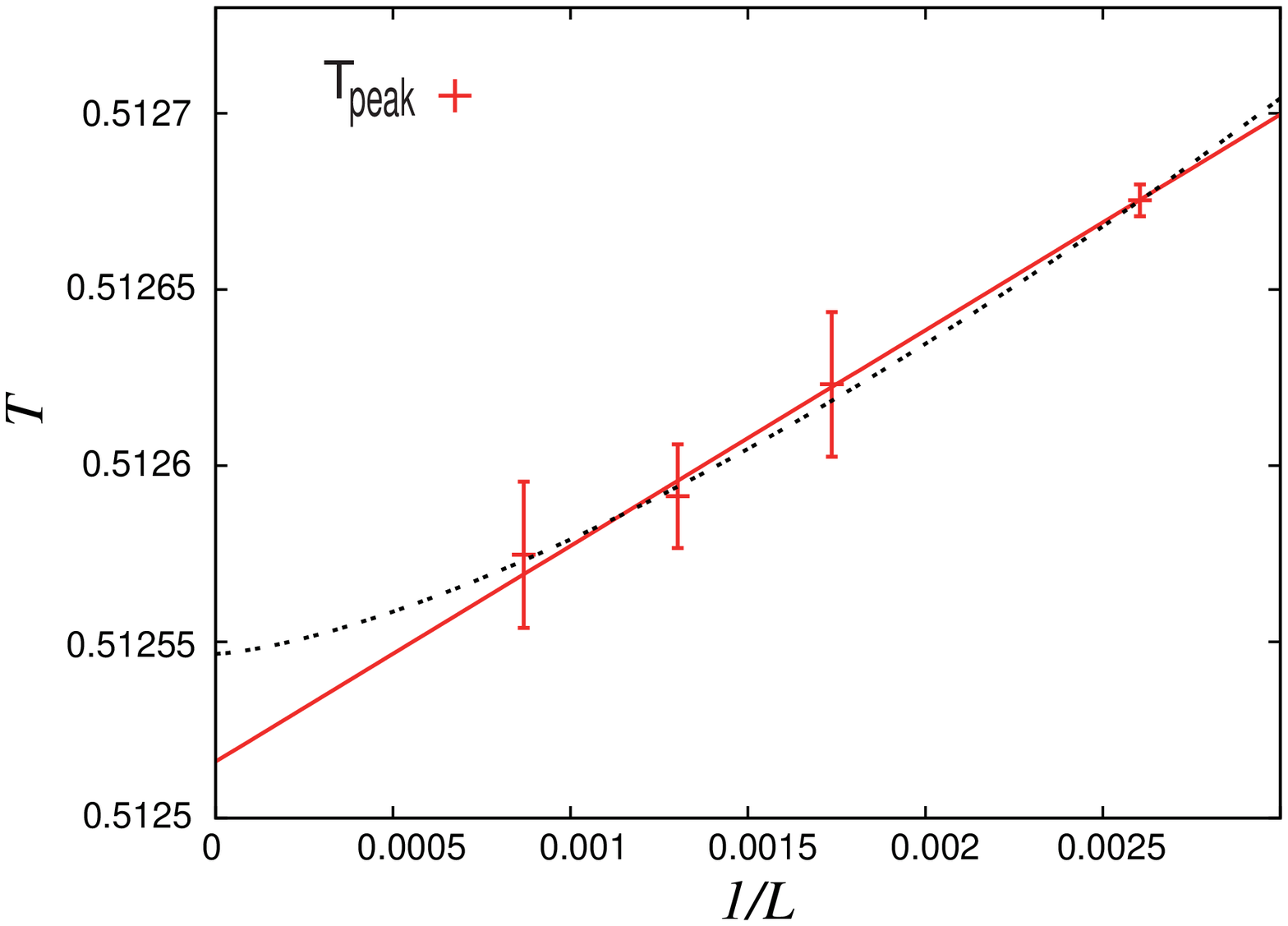}
\caption{\Lfig{peakC}(Color online) The peak temperatures of the specific heat plotted versus the inverse linear size $1/L$. The upper figure shows the data for a wide range of $L$, while the lower one is a magnified view of larger sizes $L>L_{\times}\approx 300$. The dotted straight line (blue) is a linear fit of the data for $L<300$, and the solid straight line (red) is the one for $L>300$. The dotted curve (black) represents a nonlinear fit of the data of $L>300$ on the basis of \Req{T_peak} with $\nu_c$ a fitting parameter.
}
\end{figure}

 In \Rfig{gzi_s-Tc}, we plot  the finite-size spin correlation length $\xi_s(L)$ versus the system size $L$ at a temperature $T=0.51251$ which is close to our estimate of $T_c$ (see below). One sees from the figure that the bulk spin-correlation length at the chiral transition temperature is about $\xi_s \approx 100$, which is comparable to the observed crossover size $L_{\times}\approx 300$. Then, a natural explanation of the crossover behavior observed at $L=L_{\times}$ might be the ``spin-chirality decoupling'' phenomenon occurring at longer length scales. Namely, the chiral correlation length outgrows the spin correlation length in the vicinity of $T_c$ at around the length scale $L_{\times}$.  At shorter length scales or for smaller sizes, the spin and the chirality are trivially coupled, and the spin correlation apparently dominates the ordering. At longer length scale or for larger sizes, the chiral correlation exceeds the spin one around $T_c$, and eventually dominates the ordering of the system. The observed size crossover might be regarded as a manifestation of this crossover from the spin-chirality coupling behavior for smaller sizes to the spin-chirality decoupling behavior for larger sizes. This means that the asymptotic critical behavior is expected only for larger systems of $L>L_{\times}\approx 300$.

\begin{figure}[tbp]
\includegraphics[width=1\columnwidth,height=0.7\columnwidth]{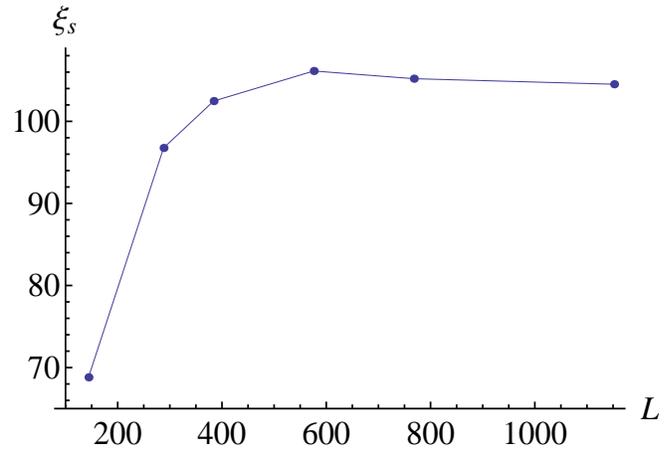}
\caption{\Lfig{gzi_s-Tc}(Color online) The size dependence of the finite-size spin correlation length at $T=0.51251\simeq T_c$. The value converges to $\xi_s \approx 100$ for large enough $L$.}
\end{figure}

 In view of this, we extrapolate the peak temperatures of the specific heat by using the data of $L>L_{\times}$ only, on the basis of the asymptotic form,
\be 
T_{\rm peak}(L)=T_{c}+({\rm const})L^{-1/\nu_{c}}, 
\Leq{T_peak}
\ee
where $\nu_c$ is the chiral correlation-length exponent.  The near-linear dependence observed in the lower panel of \Rfig{peakC} implies $\nu_{c}\simeq 1$, suggesting the standard 2D Ising universality for the chiral transition. The straight-line fit with $\nu_c=1$ yields the chiral transition temperature $T_{c-{\rm Is}}=0.51251(1)$, as indicated by the straight line (red) in the lower panel of \Rfig{peakC}. If, on the other hand, we do not fix $\nu_c$ and leave it as a free parameter in the fit of \Req{T_peak}, we get $T_{c}=0.51254(3)$ and $\nu_{c}=0.72(25)$, which is represented by the dotted curve (black) in the lower panel of \Rfig{peakC}. The quoted error bars are estimated by the standard $\chi$-square method. In fact, the resulting value of $\nu_c$ is still compatible with the Ising value $\nu_c=1$ within the error margin. 

 Next, we analyze the temperature and size dependence of the chiral Binder parameter $g_{c}$ and the chiral correlation-length ratio $\xi_{c}/L$, which are shown in \Rfig{g_c} and \Rfig{gzi_c}, respectively.
\begin{figure}[htbp]
\includegraphics[width=1\columnwidth,height=0.7\columnwidth]{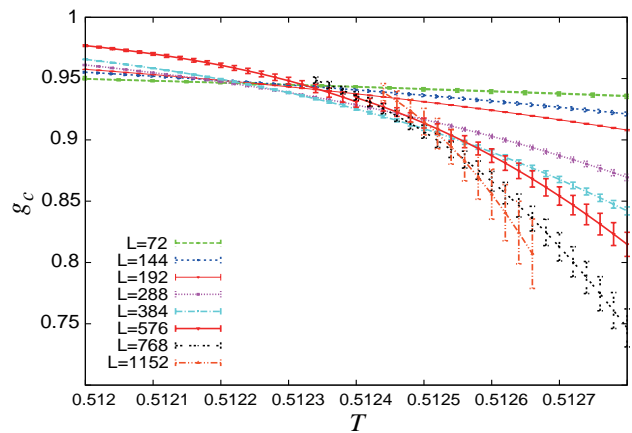}
\caption{\Lfig{g_c}(Color online) The temperature dependence of the chiral Binder parameter $g_{c}$.}
\end{figure}
\begin{figure}[htbp]
\includegraphics[width=1\columnwidth,height=0.7\columnwidth]{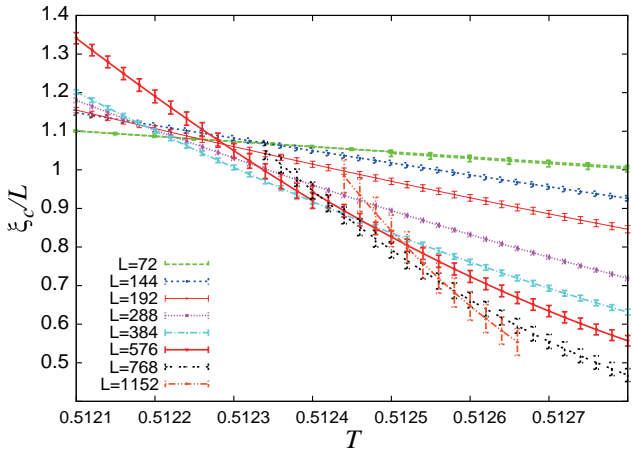}
\caption{\Lfig{gzi_c}(Color online) The temperature dependence of the chiral correlation-length ratio $\xi_{c}/L$.}
\end{figure}
Both quantities exhibit the crossing behavior around the chiral transition temperature as expected. The size dependence of the crossing temperatures between the sizes $L$ and $2L$, $T_\times(L)$, are shown in \Rfig{T_c}. 
\begin{figure}[h]
\includegraphics[width=0.92\columnwidth,height=0.68\columnwidth]{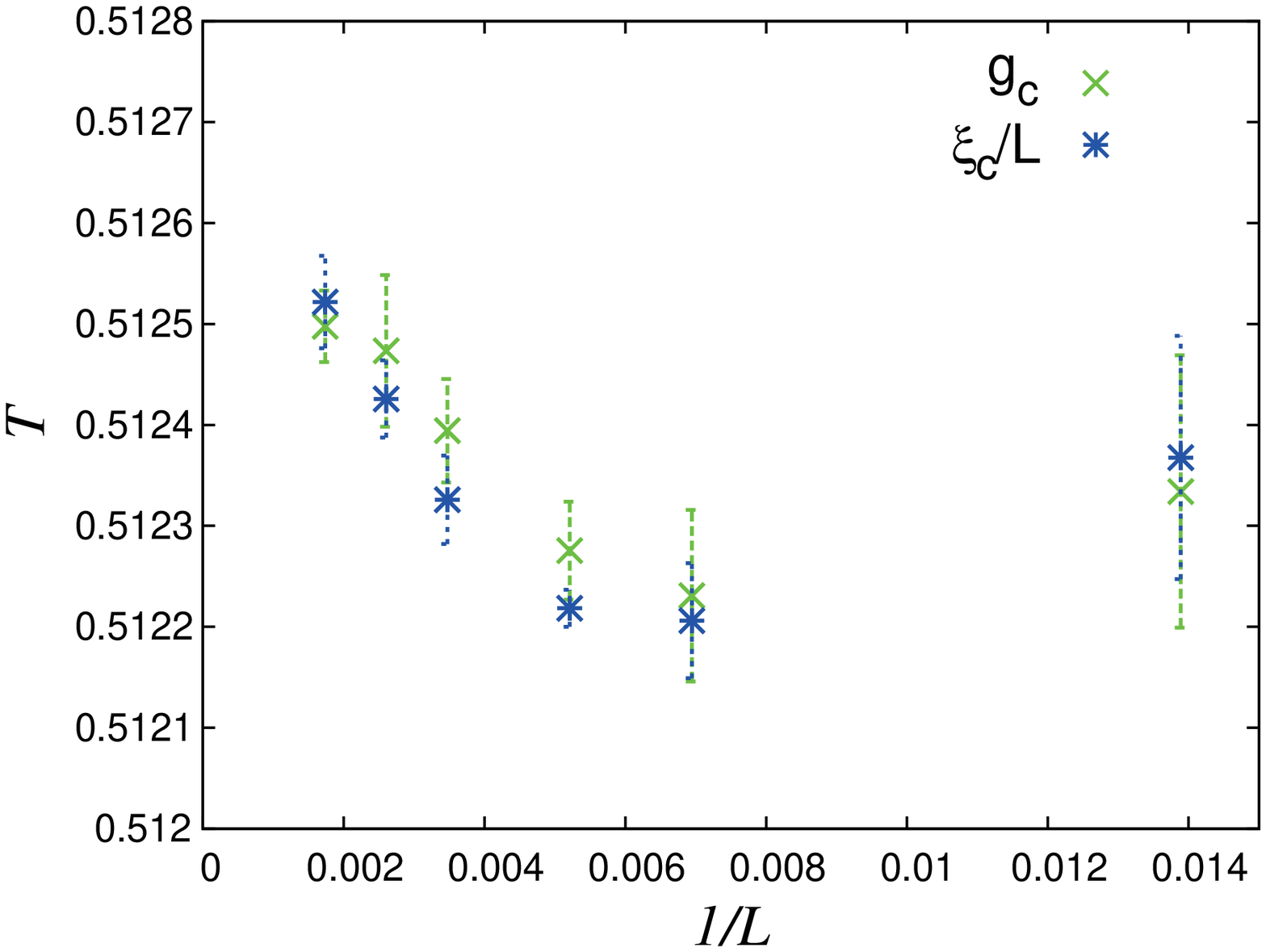}
\includegraphics[width=0.92\columnwidth,height=0.68\columnwidth]{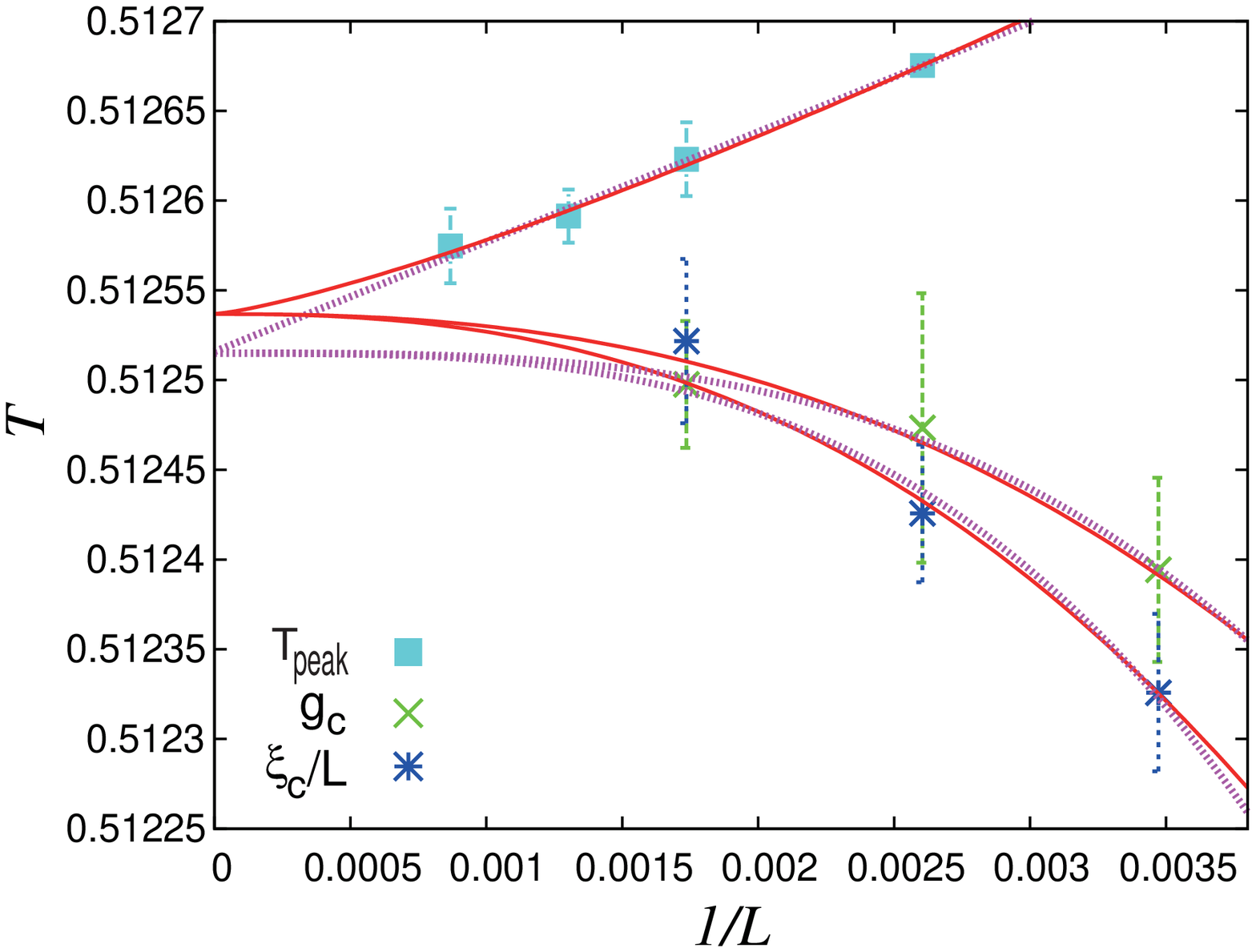}
\includegraphics[width=0.94\columnwidth,height=0.68\columnwidth]{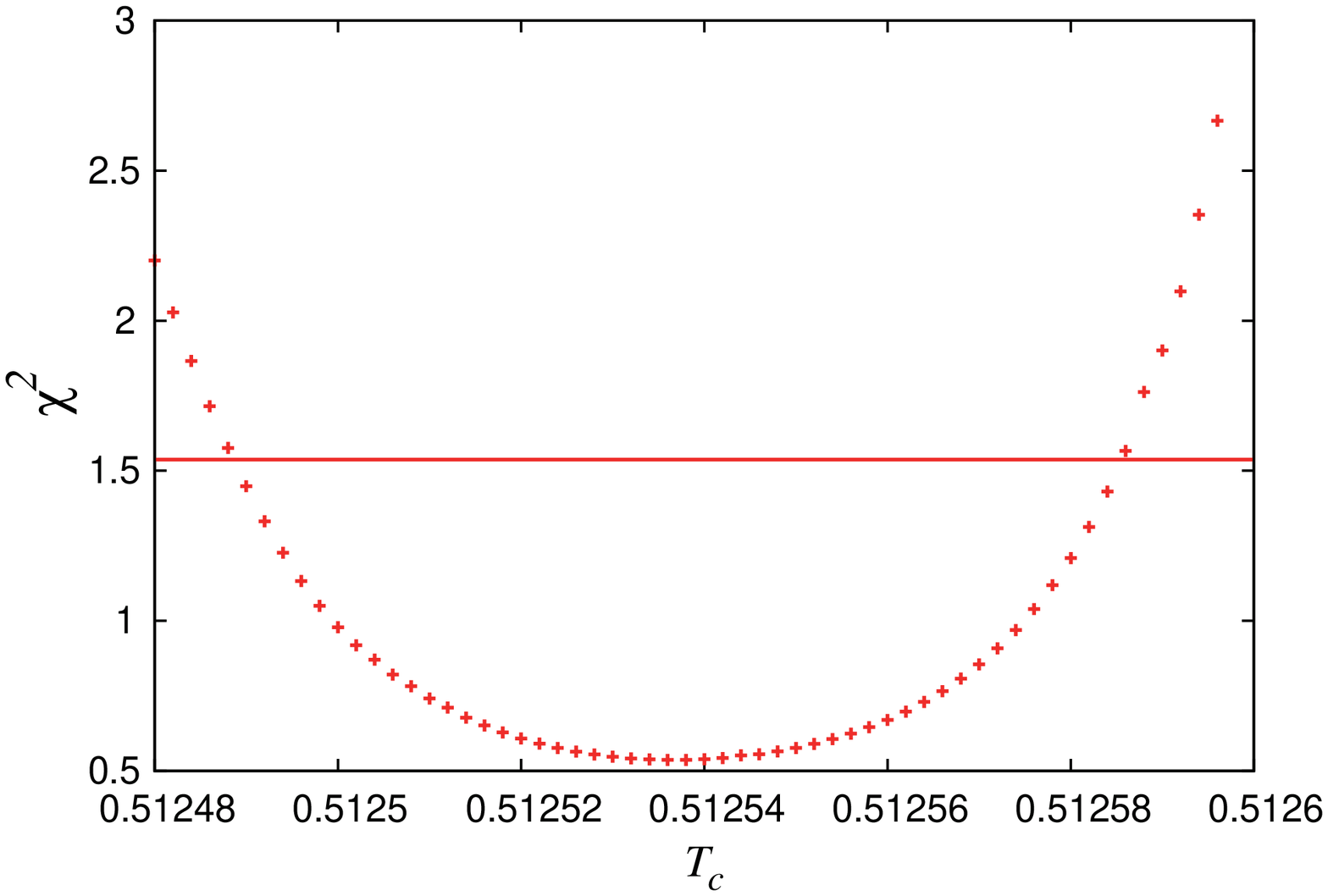}
\caption{\Lfig{T_c}(Color online) (Upper)  The inverse size dependence of the crossing temperatures of the chiral Binder parameter and of the chiral correlation-length ratio. The crossing temperatures are the ones between sizes $L$ and $2L$. (Middle) A magnified view of the upper figure in the large-$L$ region. The straight line (magenta) gives the fit of the specific-heat-peak temperatures with the Ising value $\nu_c=1$, which yields $T_c=0.51251$. The broken curves (magenta) are the combined fits of the chiral Binder parameter and of the chiral correlation-length derived under those $T_c$ and $\nu_c$. The solid curves (red) are the unbiased combined fit of all three quantities without fixing any parameters. (Lower) The total $\chi$-square values of the combined fit of the three quantities plotted against the transition temperature $T_c$ assumed in the fit. The horizontal line represents the total $\chi$-square value which is greater than the optimal value by unity, usually used as an error criterion.}
\end{figure}
Interestingly, the size dependence of $T_\times$ changes drastically around the crossover size $L_\times \simeq 300$ as already observed above for the specific-heat-peak temperatures. Namely, for $L\lesssim L_\times$, $T_\times$ tends to {\it decrease\/} with increasing $L$, while, for $L\gtrsim L_\times$, it tends to {\it increase\/} with $L$. Asymptotic behavior is then expected only for larger sizes $L\gtrsim L_\times$. We fit $T_\times$ of both the chiral Binder parameter and the chiral correlation-length ratio in the range of $L\geq 384$ to the standard finite-size-scaling form with the correction term, 
\be
T_{\times}(L)=T_{c}+({\rm const})L^{-(\nu_c^{-1} + \omega)},
\Leq{T_c_cross}
\ee
where $\omega$ is the leading correction-to-scaling exponent. The exponent $\omega$ should be common for both the Binder parameter and the correlation-length ratio, while the constant term is not so.

 We try the combined fit of all the three quantities investigated, {\it i.e.\/}, the specific-heat peak temperatures, the crossing temperatures of the chiral Binder parameter, and those of the the chiral correlation-length ratio, with $T_c,\nu_c$, and $\omega_c$ as fitting parameters. The result is $T_c=0.51254(3),\, \nu_{c}=0.79(23)$, and $\omega=1.2(11)$, which is our best unbiased estimate of $T_c,\nu_c$ and $\omega_c$. The associated fitting curve is given in  the middle panel of \Rfig{T_c} (the solid curve in red).  In the lower panel, we plot the total $\chi$-square values of this fit against the $T_c$-value assumed in the fit.  If we instead fix the values of $T_c$ and $\nu_c$ to the Ising values as determined above from the specific heat, $T_{c-{\rm Is}}=0.51251$ and $\nu_c=1$, the fit yields $\omega=2.2(6)$, which is well consistent with the Ising value $\omega=2$. The associated fitting curve is also given in the middle panel of \Rfig{T_c} (the broken curve in magenta).

 For further information about the criticality of the chiral transition, we analyze the chiral susceptibility $\chi_{c}$. In \Rfig{chi_c}, we show the size dependence of $\chi_c$ for several temperatures around $T_c$. At the chiral transition temperature $T_c$, we expect a power-law behavior $\chi_{c}\propto L^{2-\eta_{c}}$ for large enough systems.
\begin{figure}[tbp]
\includegraphics[width=0.95\columnwidth,height=0.7\columnwidth]{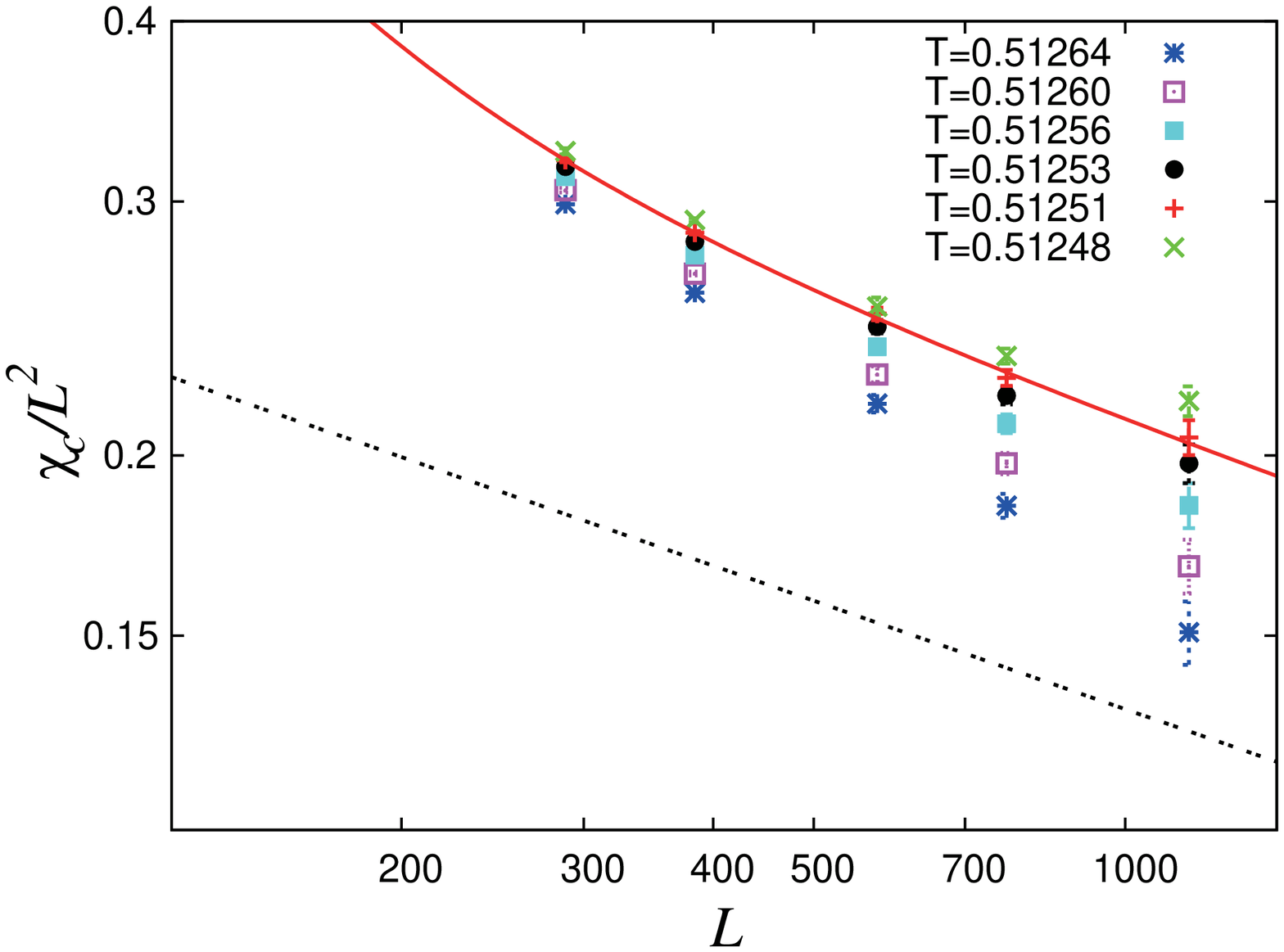}
\includegraphics[width=0.98\columnwidth,height=0.7\columnwidth]{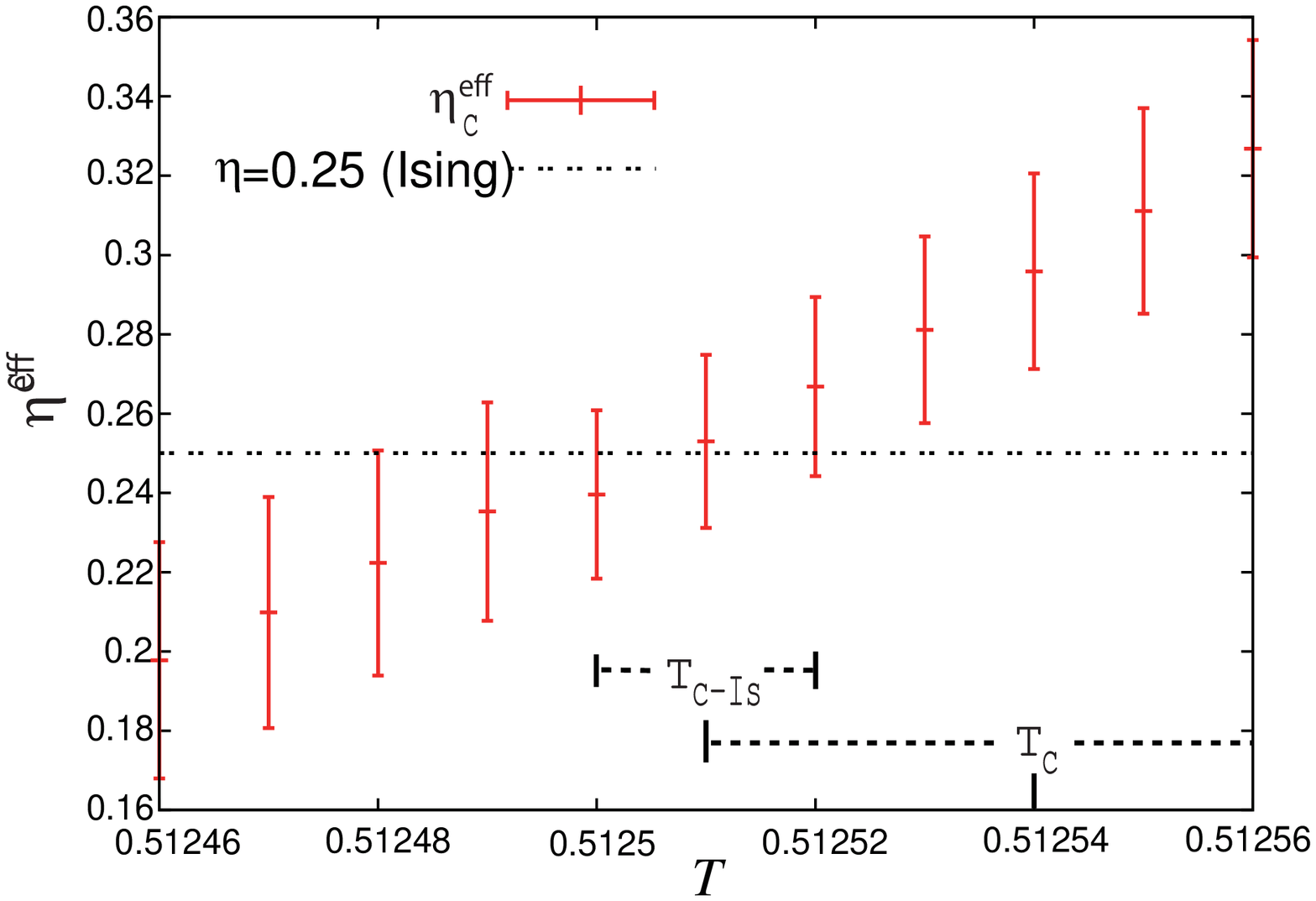}
\caption{\Lfig{chi_c}(Color online) (Upper) The chiral susceptibility divided by $L^2$, $\chi_{c}/L^2$, plotted versus the linear size $L$ at several different temperatures around $T_c$. The solid line (red) represents a fit to \Req{chi_c} at $T=0.51251$. The dotted line (black) expresses $a \times L^{-\eta}$ with the Ising exponent $\eta=0.25$, and is plotted as a guide to the eye. (Lower) The temperature dependence of the effective chiral anomalous-dimension exponent $\eta_{c}^{\rm eff}(T)$ evaluated through \Req{chi_c}. The estimated transition temperatures $T_{c-{\rm Is}}=0.51251(1)$ and $T_c=0.51254(3)$ (see the main text) are given with the error bars. The Ising value $\eta=1/4$ is obtained around $T=T_{c-{\rm Is}}=0.51251$. }
\end{figure}
A complete linearity is not observed at any temperature, which implies that the scaling region has not completely been reached. Hence, we analyze the data on the basis of the following scaling form with a constant correction term,
\be
\chi_{c}(T,L)=a(T)+b(T)L^{2-\eta_c^{\rm eff}(T)}.
\Leq{chi_c}
\ee
By fitting the data for large sizes of $L\geq 288$, we obtain the effective value of the chiral anomalous-dimension exponent $\eta_c^{\rm eff}(T)$, and the results are plotted versus the assumed critical temperature in \Rfig{chi_c}. We find that the Ising value $\eta=1/4$ is obtained around $T=0.51251$, which is quite consistent with the $T_c$-value deduced from the specific-heat-peak temperatures $T_{c-{\rm Is}}=0.51251(1)$ with assuming the Ising exponent $\nu_c=1$. Meanwhile, our unbiased estimate $T_{c}=0.51254(3)$ associated with $\nu_{c}=0.79(23)$ yields somewhat larger value for $\eta_c$, $\eta^{\rm eff}_{c}=0.295(63)$, which is still compatible with the 2D Ising universality class within the error margin, though the error bar is now pretty large.

Summarizing the above results, we find that our unbiased estimates of $T_c,\nu_c$ and $\omega$ are compatible with the Ising universality, though a new universality class is also admissible within, say, 20\% accuracy for the exponents. If we presume the Ising value $\nu_c=1$, we can precisely locate the transition temperature $T_{c}=0.51251(1)$. This value of $T_{c}$ shows a good agreement with the $T_c$-value derived from the chiral susceptibility under the Ising assumption $\eta=1/4$. Such consistency certainly favors the Ising universality for the chiral transition, though further investigations are desirable to completely settle the issue.

\subsection{Spin transition}

Next, we move to the spin ordering at $T=T_s$. The spin Binder parameter $g_{s}(T,L)$ and the spin-correlation-length ratio $\xi_{s}(T,L)/L$ of the sizes $L=36,72,144,288,576,1152$ are shown in \Rfig{g_s} and \Rfig{gzi_s}, respectively. Other sizes $L=48,96,192,384,768$ are also computed, but we omit them here for clarity since they show a similar behavior.
In \Rfig{v}, we also plot the reduced vorticity modulus $v(T|2L,L)$ versus the temperature for $L=36,72,144,288,576$. 
\begin{figure}[htbp]
\includegraphics[width=1\columnwidth,height=0.7\columnwidth]{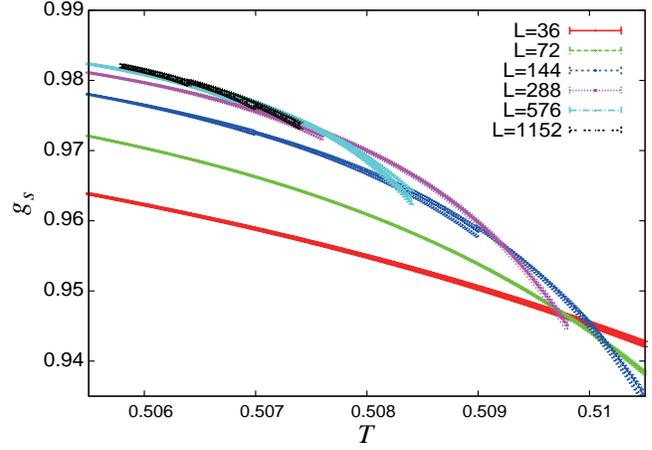}
\caption{\Lfig{g_s}(Color online) The temperature dependence of the spin Binder parameter $g_{s}$ for system sizes $L=36, 72, 144, 288, 576, 1152$.}
\end{figure}
\begin{figure}[htbp]
\includegraphics[width=1\columnwidth,height=0.7\columnwidth]{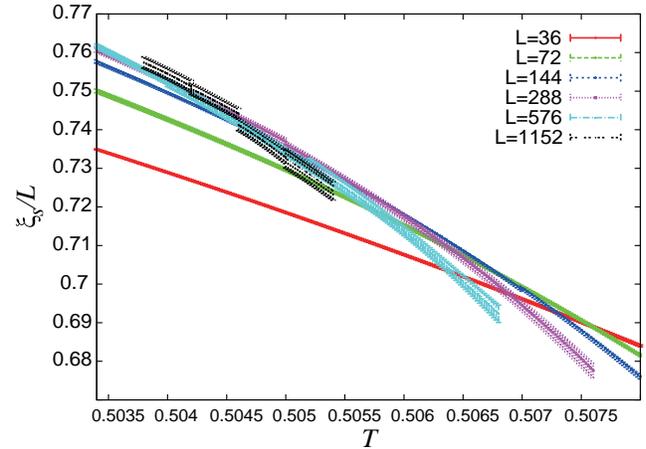}
\caption{\Lfig{gzi_s}(Color online) The temperature dependence of the spin  correlation-length ratio $\xi_{s}/L$ for system sizes $L=36,72,144,288,576,1152$.}
\end{figure}
\begin{figure}[htbp]
\includegraphics[width=1\columnwidth,height=0.7\columnwidth]{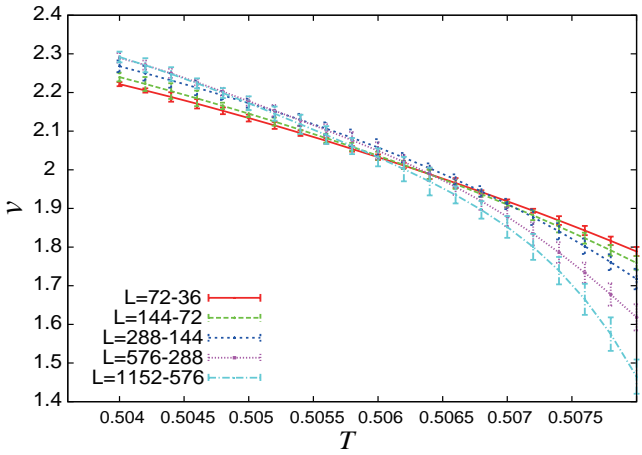}
\caption{\Lfig{v}(Color online) The temperature dependence of reduced vorticity modulus $v(T|2L,L)$ with $L=36,72,144,288,576$.}
\end{figure}

 From \Rfigss{g_s}{v}, we can see immediately that the crossing temperatures for the spin related quantities are located in the temperature range significantly lower than the chiral transition temperature estimated above, $T_c \simeq 0.5125$, thereby providing further evidence that the spin transition occurs at a temperature lower than the chiral transition, {\it i.e.\/}, the the spin-chirality decoupling. We also see from these figures that the data at lower temperatures tend to exhibit a weaker and weaker splay-out behavior for larger sizes, and instead show a merging behavior. Such a merging behavior, typically observed in systems showing the KT transition, is certainly expected for the present system exhibiting a spin quasi-long-range order.

 The size dependence of the crossing temperatures between the sizes $L$ and $2L$ are summarized in \Rfig{T_s}. To estimate the spin transition $T_s$, we fit the $L$-dependence of the crossing temperatures $T_{\times}(L)$ to the form,
\be
T_{\times}(L)=T_{s}+{\rm (const)}(\ln L)^{-x}.
\Leq{T_s}
\ee
The value of $T_s$ and the exponent $x$ should be common for all the three quantities, while the constant factor is not so. 
\begin{figure}[tbp]
\includegraphics[width=1\columnwidth,height=0.7\columnwidth]{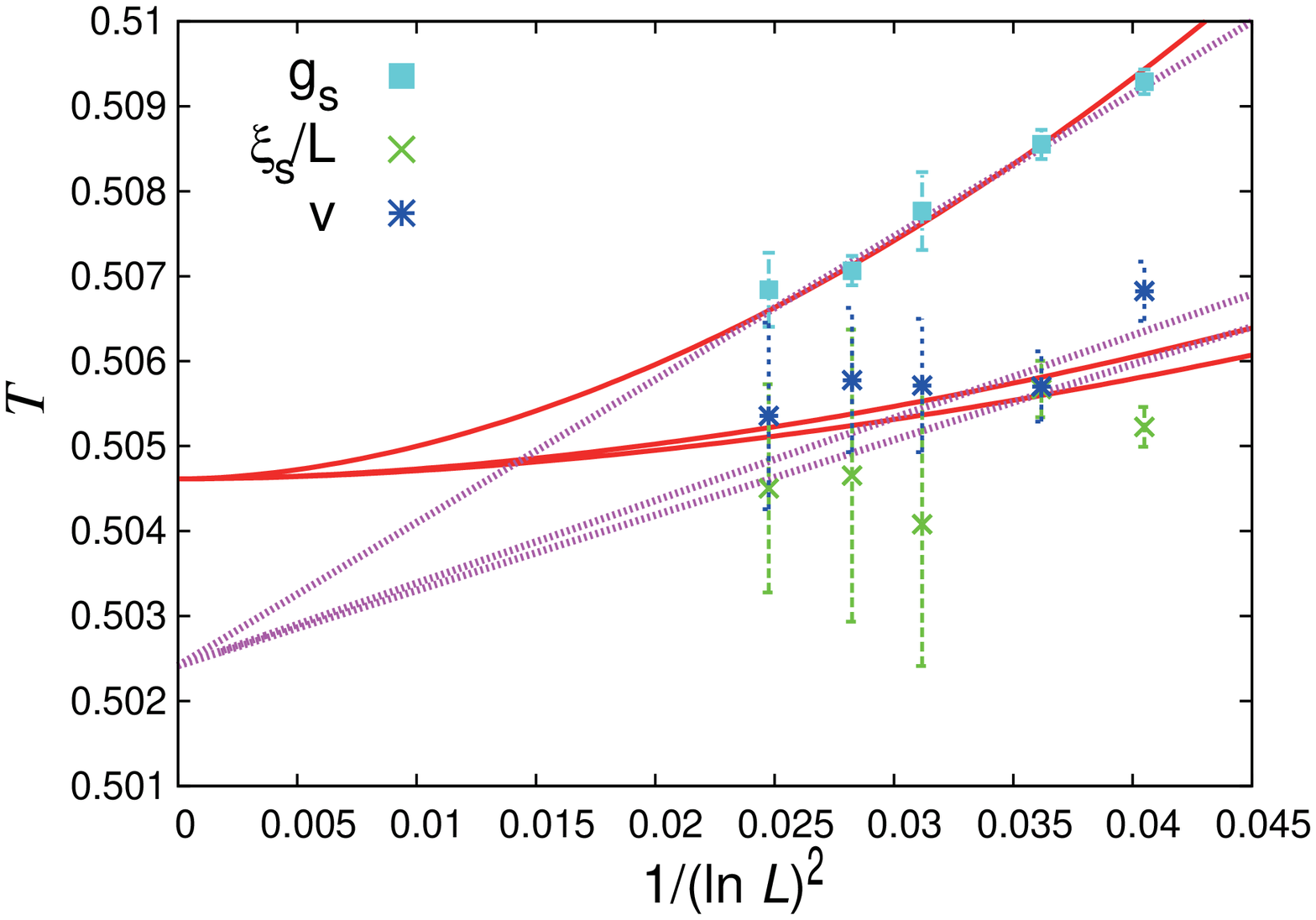}
\includegraphics[width=1\columnwidth,height=0.7\columnwidth]{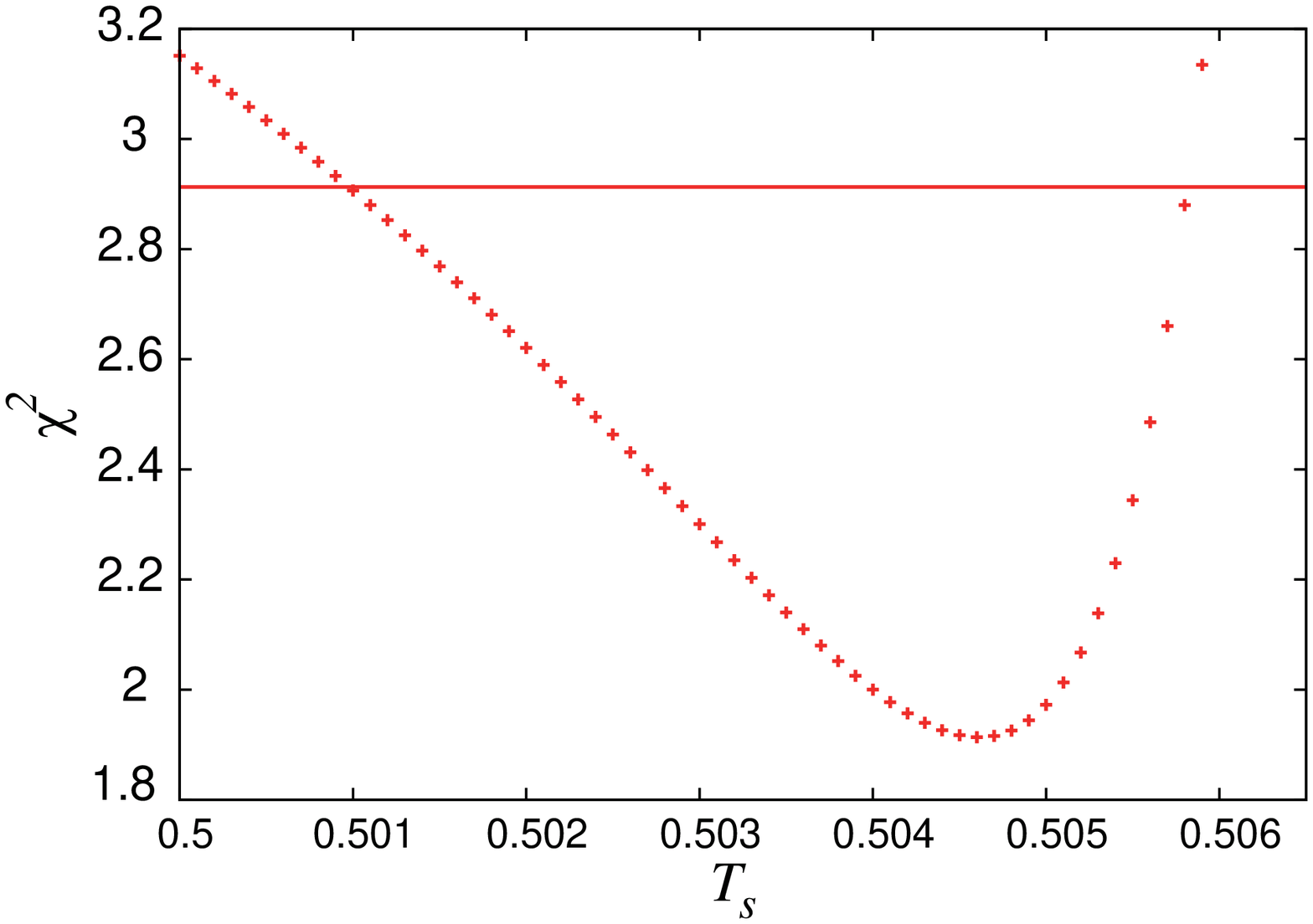}
\caption{\Lfig{T_s}(Color online) (Upper) The crossing temperatures of the spin Binder parameter $g_{s}$, of the spin correlation-length ratio $\xi_{s}/L$, and of the reduced vorticity modulus $v$ are plotted versus $(\ln L)^{-2}$. The straight (magenta) and curved (red) lines are the fits on the basis of \Req{T_s}. The straight ones correspond to the KT fit ($x=2$ in \Req{T_s}) and the curves are the non-KT fit with treating $x$ as a fitting parameter. $T_s$ and $x$ are taken to be common in each fit. (Lower) The $\chi^2$ values of the non-KT fit on the basis of (\ref{eq:T_s}) are plotted against the assumed $T_s$-values, where $x$ is adjusted as a function of $T_s$. The horizontal straight line denotes the minimum $\chi^2$-value plus unity.}
\end{figure}

We perform the two types of fits on the basis of \Req{T_s}: In one, we fix $x=2$, while, in the other, we treat $x$ as a free fitting parameter. The former corresponds to the conventional KT fit, while the latter allows for the non-KT fit. The former KT fit with $x=2$ yields $T_{s-{\rm KT}}=0.5024(5)$, while the latter fit yields $T_{s-{\rm nonKT}}=0.5046(10)$ and $x=3.6(12)$. To see the relevancy of these two types of fits, we plot the associated $\chi^2$-values of the fits as a function of the assumed $T_s$-value in the lower panel of \Rfig{T_s}, where the optimal $x$ has been adjusted as a function of $T_s$. As can be seen from the figure, the best fit is obtained for the non-KT fit with $T_{s-{\rm nonKT}}=0.5046$, but the KT-fit with $T_{s-{\rm KT}}=0.5024$ lies within the error margin due to the asymmetry of the $\chi^2$-curve, and cannot be ruled out. 

 In order to discriminate between the KT and the non-KT criticalities for the spin transition, we proceed to examine other quantities. We first evaluate the spin anomalous-dimension exponent $\eta_s$ by analyzing the $K$-point spin susceptibility $\chi_s$. At and below the spin transition temperature $T_s$, a power-law behavior $\chi_{s}(T,L)\propto L^{2-\eta_s(T)}$ is expected. In \Rfig{chi_s}, we plot the spin susceptibility $\chi_{s}$ divided by $L^2$ as a function of $L$ on a log-log plot, and we can see a good linearity of the data, in which the scaling region is expected. Thus, the effective spin anomalous-dimension exponent $\eta_{s}^{\rm eff}(T)$ can be defined through the form
\be
\chi_{s}(T,L)=a(T)L^{2-\eta_{s}^{\rm eff}(T)}.
\Leq{chi_s}
\ee
The estimated $\eta_{s}^{\rm eff}(T)$ is plotted in the lower panel of \Rfig{chi_s}.
\begin{figure}[tbp]
\includegraphics[width=0.95\columnwidth,height=0.7\columnwidth]{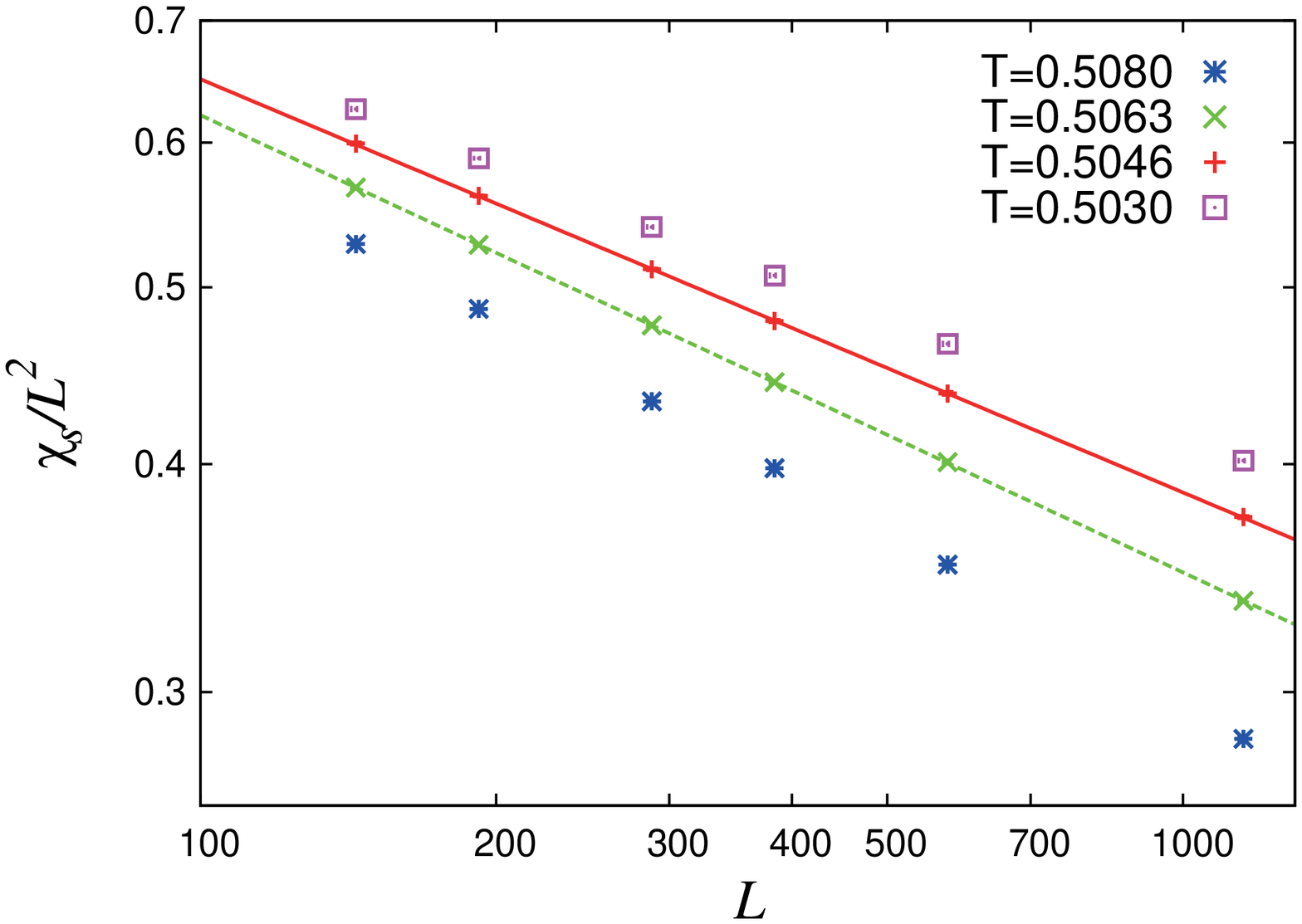}
\includegraphics[width=0.98\columnwidth,height=0.7\columnwidth]{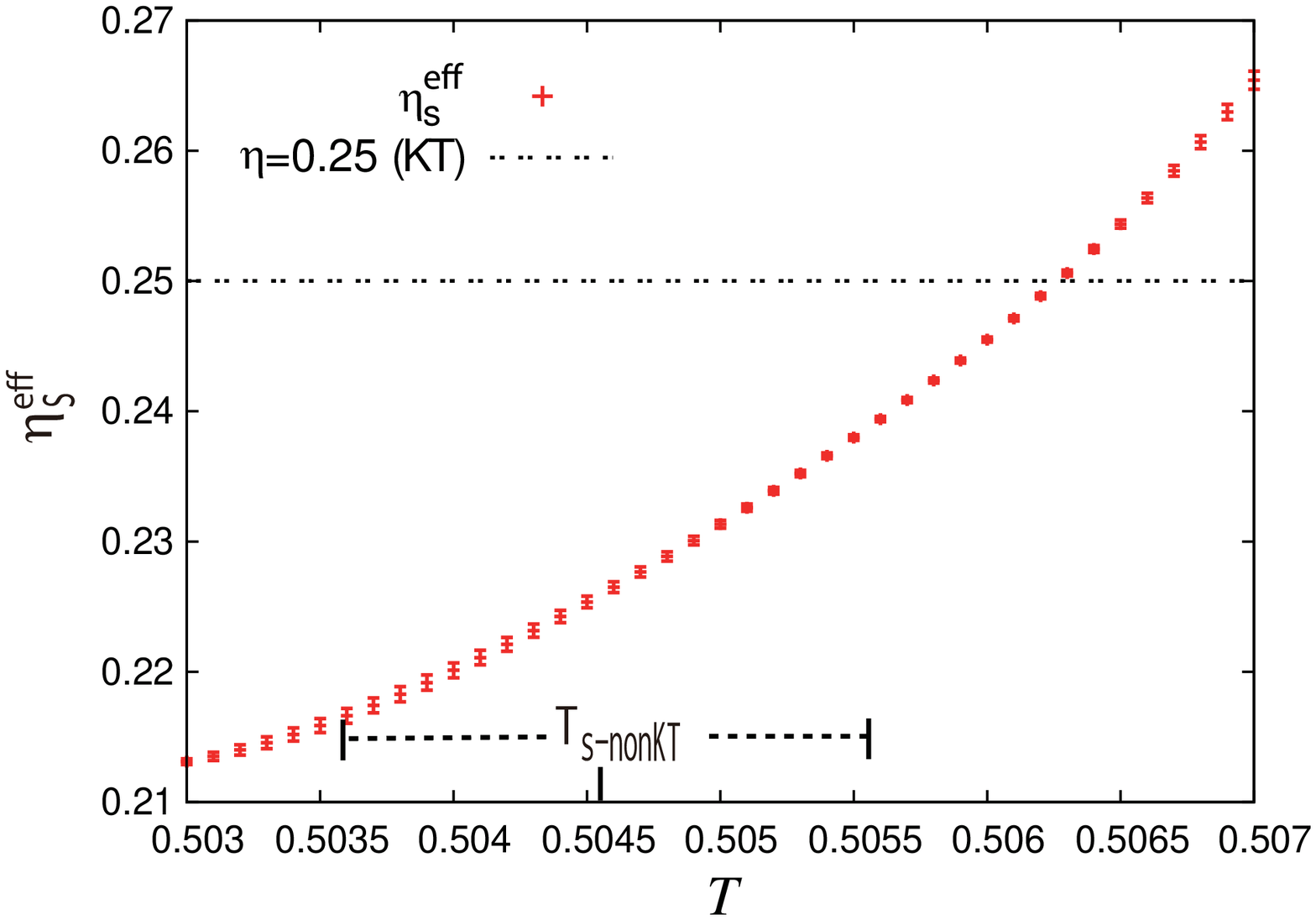}
\caption{\Lfig{chi_s}(Color online) (Upper) The system-size $L$ dependence of the spin susceptibility divided by $L^2$ at several temperatures on a log-log plot. The solid lines demonstrate the power law $L^{-\eta_s}$ with $\eta_s=0.25$ and $\eta_s=0.226$. The conventional KT value $\eta_s=0.25$ is denoted by the broken line (green) obtained by fitting the data at $T=0.5063$,  the temperature associated with the KT transition, while the solid line (red) represents $\eta_s=0.226$ obtained by fitting the data at $T=0.5046$, the temperature associated with the non-KT transition (see the main text). Note that the $\chi_s(T)$ data of $L=768$ and $1152$ in a temperature range $T<0.5038$ is obtained by a polynomial extrapolation of $\chi_s(T)$ at higher temperatures.
(Lower) The effective spin anomalous-dimension exponent $\eta_s$ plotted versus the assumed $T_s$-value. The estimated transition temperature $T_{s-{\rm nonKT}}=0.5046(10)$ is also given with the error bar. The KT transition temperature $T_{s-{\rm KT} }=0.5024(5)$ is located outside this temperature range.
}\end{figure}
From the figure,  we find $\eta_s=0.226(14)$ at $T_{s-{\rm nonKT}}=0.5046$. If we assume the KT value of $\eta_s=0.25$, we find $T_s=0.5063(1)$, which happens to be significantly higher than $T_{s-{\rm KT}}=0.5024(5)$ estimated above from the crossing temperatures of the Binder ratio, of the correlation-length ratio and of the vorticity modulus. This discrepancy indicates that the KT criticality $\eta_s=0.25$ and $x=2$ might be inappropriate in describing the spin transition of the present model.

 We also employ the helicity modulus to further examine the consistency. In 2D {\it XY\/} models, the helicity modulus is expected to show a discontinuous jump of magnitude $T_{s}/(2\pi \eta_s(T_{s}))$ at the transition temperature $T_s$. This quantity relates the transition temperature $T_s$ to the exponent $\eta_s$.  The temperature dependence of the helicity modulus is shown in \Rfig{Y}. In the figure, we show the two straight lines $T/(2\pi \eta_s)$, one with the KT value of $\eta_s=1/4$ and the other with the non-KT value $\eta_s=0.226$ as estimated above. By extrapolating the crossing temperatures between these two straight lines and  the helicity-modulus data to $L=\infty$ on the basis of (\ref{eq:T_s}), we can obtain an estimate of $T_s$. The resulting extrapolation curves are also given in \Rfig{Y}.
\begin{figure}[tbp]
\includegraphics[width=1\columnwidth,height=0.7\columnwidth]{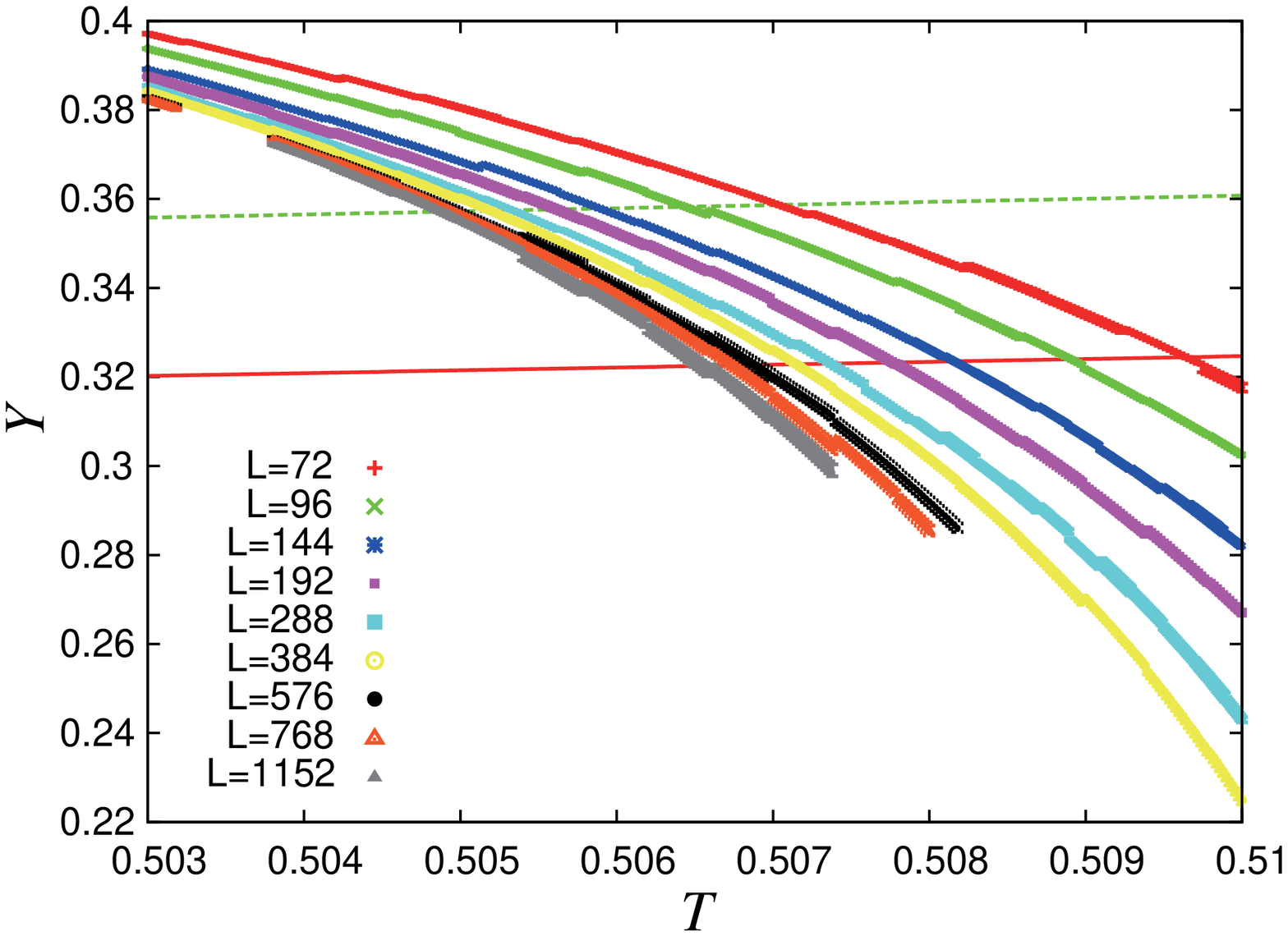}
\includegraphics[width=1\columnwidth,height=0.7\columnwidth]{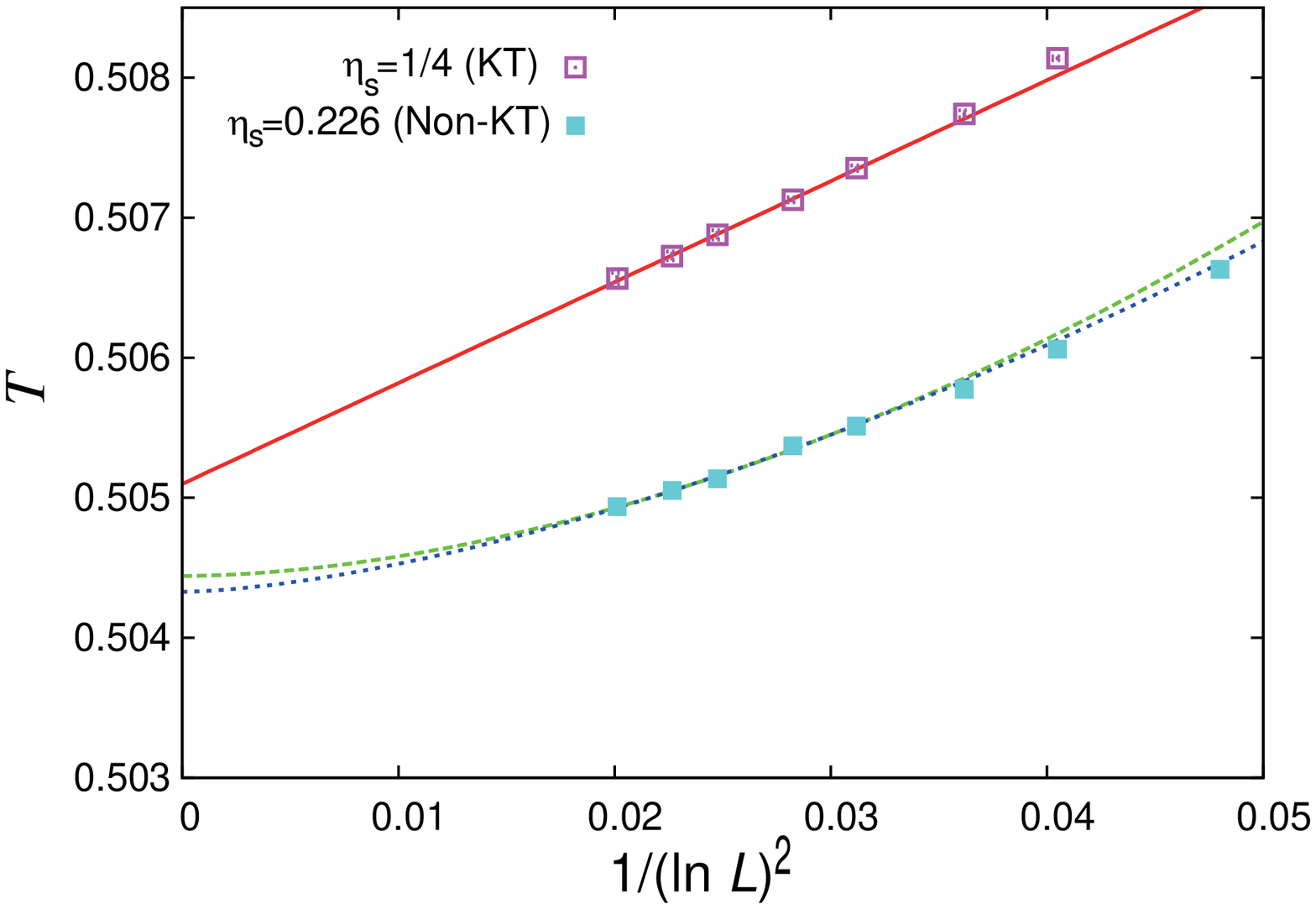}
\caption{\Lfig{Y}(Color online) (Upper) The temperature dependence of the helicity modulus of different sizes. The lower solid straight line (red) and the upper broken straight line (green) denote $T/(2\pi \eta_s)$ with $\eta_s=1/4$ and $\eta_s=0.226$, respectively. (Lower) The intersection points of the helicity-modulus data and the straight lines of the upper figure are plotted against $(\ln L)^{-2}$. The solid line (red) is the fit under the KT assumption $\eta_s=1/4,x=2$, and the broken line (green) is the fit under the non-KT assumption $\eta_s=0.226,x=3.6$. The former KT fit yields $T_s=0.5051(1)$, whereas the latter non-KT fit yields $T_s=0.5044(1)$. The broken line (blue) is the fit with $\eta_s=0.226$ and with $x$ as a free fitting parameter.}
\end{figure}
 For the KT case $\eta_s=1/4$ and $x=2$, we find $T_{s}=0.5051(1)$, which, however, turns out to be inconsistent again with the $T_s$-values estimated above by other independent methods under the KT assumption, {\it i.e.\/}, $T_s=0.5024(5)$ and $T_s=0.5063(1)$. On the other hand, for the non-KT case $\eta_s=0.226$ and $x=3.6$, we get $T_{s}=0.5044(1)$ which turns out to accord well with the $T_s$-value, $T_{s-{\rm nonKT}}=0.5046(10)$, estimated above by the other methods. We also try the fit where $x$ is treated as a free fitting parameter. The result is $T_s=0.5043(1)$ and $x=3.2(18)$, which yields essentially the same $T_s$ as the constrained non-KT fit.

 Considering all these results, we may now exclude the standard KT criticality for the spin transition. The spin transition of the present model has a non-KT character. Our best unbiased estimates of the spin-transition temperature and the spin anomalous dimension are $T_s=0.5046(10)$ and  $\eta_s=0.226(14)$, respectively. We note that the possibility of the non-KT nature of the spin transition was suggested before by several authors~\cite{Teitel:83,S_Lee:98}. Here we demonstrate it with a higher confidence level by performing larger-scale simulations. We should also mention that the recent result on the square-lattice FFXY model by Okumura {\it et al\/}~\cite{Okumura:11} gave essentially the same conclusion. The observed consistency between the square and triangular lattices may suggest a generality of the non-KT nature of the spin transition in the 2D FFXY models.

\section{\Lsec{summary}Summary and Discussions}

In this paper, we investigated the AF {\it XY\/} model on the triangular lattice by extensive MC simulations, focusing on its spin and chirality orderings. Large systems up to $L=1152$ were simulated. Our data unambiguously showed that the chiral and spin transitions took place at two different temperatures, {\it i.e.\/} the occurrence of the spin-chirality decoupling. The spin transition temperature turned out to be lower than the chiral transition temperature by about 1.5\%. A crossover behavior from the spin-chirality coupling regime to the decoupling regime was observed around the size $L_{\times}\approx 300$. This crossover is associated with the interchange of the relative dominance of the spin and the chiral correlation lengths. 

The universality class of the chiral and the spin transitions was also examined. Concerning the criticality of the chiral transition, although we cannot completely rule out the possibility of the non-Ising universality class, the high level of consistency we observed among different independent observables certainly speaks for the Ising universality class. If we presume the Ising nature of the chiral transition, {\it i.e.\/}, presume $\nu_c=1$, the chiral transition temperature was estimated quite accurately as $T_{c}=0.51251(1)$. If we do not assume the Ising nature of the chiral transition, we get $T_c=0.51254(3)$, $\nu_c=0.79(23)$ and $\eta_c=0.295(63)$ which is still compatible with the Ising criticality, though the error bars are pretty large. 

 Let us now compare our present estimates of $T_c$, $\nu_c$, and $\eta_c$ with the recent estimates by other authors for the same model~\cite{S_Lee:98,Ozeki:03}. On the basis of an equilibrium MC simulation, S. Lee and K.-C. Lee gave $T_{c}=0.513(1)$, $\nu_c=0.833(8)$ and $\eta_c=0.25(3)$~\cite{S_Lee:98}, which accord with ours. However, their data were taken only for rather small sizes of up to $L=108$, smaller than the crossover length scale $L_{\times}\sim 300$, so that the data might fail to capture the genuine scaling behavior. As such, the accordance of their results with our present estimates may partly be accidental. On the basis of the nonequilibrium relaxation method, Ozeki and Ito gave  $T_c=0.512(1)$, $\nu_c=0.84(2)$ and $\eta_c=0.25(1)$, claiming  a new universality class for the chiral transition~\cite{Ozeki:03}. The method employed, {\it i.e.\/}, the nonequilibrium relaxation method, however, has an intrinsic limitation of the finite observation time in spite of the apparently large system size studied. 

 For the FFXY model on the square lattice, a model different from the present one but believed to lie in a common universality class, two extensive equilibrium MC simulations comparable in their size to our present one, have recently been made. Hasenbusch {\it et al.} studied the sizes up to $L=1000$, and concluded that the chiral transition belonged to the 2D Ising universality class~\cite{Hasenbusch:05}. More precisely, these authors estimated the effective critical exponents for finite-size systems and observed a tendency converging to the Ising criticality for larger sizes of $L \gsim 500$. Interestingly, a qualitative change of the size dependence was observed around the crossover size $L_\times \sim 500$, which accords with our present observation for the triangular {\it XY\/} model. Okumura {\it et al.} performed a large-scale MC simulation of the same model up to the size $L=1024$~\cite{Okumura:11}. These authors also concluded the Ising universality for the chiral transition, corroborating the conclusion of Hasenbusch {\it et al.}. Thus, there now seems to be a reasonable amount of numerical evidence that the chiral transition of the 2D FFXY model lies in the standard 2D Ising universality class.

%We also note that the precisions of our $\nu_c$ and $\eta_c$ are not improved in comparison with theirs, in spite of larger-scale simulations. This is again due to the crossover from the spin-chirality coupling to decoupling regimes. The finite-size-scaling scheme works only in the decoupling regime $L>300$, which makes the analysis difficult.

 Concerning the criticality of the spin transition, we observed an unacceptable level of inconsistency between several independent quantities studied when we assumed the standard KT universality class for the spin transition. Such an inconsistency was resolved once we assumed the non-KT critical exponents for the spin transition. Then, we concluded that the spin transition of the present model belonged to the non-KT universality class. As our final estimates, we quote $T_s=0.5046(10)$ and $\eta_{s}=0.226(14)$. 

 Comparing our present estimates of $T_s$ and $\eta_s$ with the recent estimates by other authors on the same model, we find that our values accord with neither $T_s=0.501(1)$ and $\eta_s=0.22(1)$ by Lee {\it et al.}~\cite{footnote1}, nor $T_s=0.508(1)$ by Ozeki {\it et al.}~\cite{Ozeki:03}. Concerning the former case, the deviation of $T_s$ is presumably due to the smallness of the simulation sizes in ref. 20. In view of the discrepancy in the estimated $T_s$-values, however, the agreement in the estimated $\eta_s$-values might probably be accidental. Concerning the latter work by Ozeki {\it et al\/}, the cause of the discrepancy might come from the tendency that the short-time observation regime usually employed in the nonequilibrium relaxation method overestimates the transition temperature.

 We also compare our estimate of $\eta_s$ with the recent estimates by Hasenbusch {\it et al.}~\cite{Hasenbusch:05} and by Okumura {\it et al.}~\cite{Okumura:11} for the square-lattice FFXY model. Hasenbusch {\it et al.} claimed the KT nature of the spin transition, based on their observation that the two independent quantities, {\it i.e.\/}, the helicity modulus and the spin correlation-length ratio, gave consistent results if one assumed the KT nature of the transition. Here we wish to comment that the type of their analysis was only a consistency check, which in itself is not enough to exclude the possible non-KT character of the transition. Okumura {\it et al.} examined the possibility of the non-KT nature of the spin transition on the same model, and reported the $\eta_s$-value as $\eta_s=0.20(1)$, which was slightly smaller than our present estimate $\eta_s=0.226(14)$, though they come near the quoted error margin. The difference of $\eta_s$ between the two FFXY models, if it really occurs, would imply the non-universal character of the criticality of the spin transition of the 2D FFXY models. Further study is required to fully substantiate such a possibility, however.

%Their analyses were based on the KT assumption and seemed not to examine possible deviations from the KT critical exponents. This possibility of deviations has been explored in rather earlier researches~\cite{Teitel:83,Minnhagen:85,Choi:85,Granato:86,J_-R_Lee:94,Grest:89,S_Lee:98}, whose estimates still show small deviations from ours, which is presumably due to the finite-size effect. We also refer to the most recent result of the FFXY model on the square lattice~\cite{Okumura:11}, 

%We should also note that we compared our result with a picture proposed by J. Lee et al.~\cite{J_Lee:91-1,J_Lee:91-2}. They have claimed that the chiral and spin transitions concurrently occur and are of first order. We think that the seeming first order transition suggested by them is just due to the finite-size effect, which is already pointed out and is intensively discussed by Hasenbusch et al.~\cite{Hasenbusch:05}. Our numerical data actually do not exhibit any anomaly of the energy and the specific heat around the spin-transition temperature, which strongly rules out the first-order transition.

 The Ising universality of the chiral transition suggested in this paper is seemingly quite natural, since the chirality is an Ising variable possessing a $Z_2$ symmetry. Nevertheless, we stress that the criticality of the chiral transition is still a nontrivial issue. In fact, in some systems such as vector spin glasses, critical exponents of the chiral transition are known to be quite different from the corresponding Ising values ({\it i.e.\/}, from those of the Ising spin glass)~\cite{Kawamura:10,Kawamura:01,Viet:09}. Hence, a naive symmetry consideration is not always enough in identifying the universality class. A possible cause of such a deviation from the symmetry-based prediction might be the long-range nature of the chirality-chirality interaction. In 2D {\it XY\/} models, the effective interaction between the chiralities is known to be Coulombic, {\it i.e.\/}, logarithmic. In the present 2D FFXY model, such a long-range nature of the chirality interaction apparently does not affect the universality class. This may be due to the simplicity of the relevant phase space of the present model, in contrast to that of spin glasses. It will then be interesting to clarify the conditions of the chiral transition belonging to the standard Ising universality class. 

 The universality class of the spin transition is also a quite nontrivial issue. Because of the spin-chirality decoupling, the spin transition of the present model occurs in the presence of the chiral order already established at a higher temperature $T_c$. This implies that the phase space available for the spin degrees of freedom at $T_s<T_c$ is rather restricted. Such a situation is very different from the spin transition of the conventional unfrsutrated 2D {\it XY\/} models where an entire phase space is available, and hence, the nature of the spin ordering of the current system could deviate from the conventional KT transition. To understand the universality of the transition, several theoretical approaches were already proposed, such as the Coulomb-gas representation of FFXY models~\cite{Okumura:11} and the renormalization-group analyses of the coupled XY models which are believed to be in the same universality class as the 2D FFXY models~\cite{Choi:85,Granato:86,Jeon:97}. However, a full comprehension of the phenomenon is still lacking. Another approach based on the KT renormalization group was proposed~\cite{Minnhagen:85}, which yielded the nonuniversal jump of the helicity modulus, although the connection to the 2D FFXY models is not necessarily clear. Thus, further studies are required to fully understand the nature of the criticality of the spin transition of the 2D FFXY models.

%%%%%%%%%%%%%%%%%%%%%%%%%%%%%%%%%%%%%%%%%%%%%%%%%
%Acknowledgement 
%%%%%%%%%%%%%%%%%%%%%%%%%%%%%%%%%%%%%%%%%%%%%%%%%
\begin{acknowledgment}

The authors are grateful to H. Yoshino and T. Okubo for useful discussions. This study is supported by Grant-in-Aid for Scientific Research on Priority Areas `Novel State of Matter Induced by Frustration' (19052006 and 19052008). A part of numerical calculations were performed on SX9 at the institute for solid state physics in University of Tokyo. 
\end{acknowledgment}


\begin{thebibliography}{99}
\bibitem{footnote1} We estimate this value of $\eta_s$ from the jump size of the helicity modulus given in ref. 20.
\end{thebibliography}
\end{document}